\documentclass{aip-cp}

\usepackage[numbers]{natbib}
\usepackage{rotating}
\usepackage{graphicx}
\usepackage{combelow}
\usepackage{braket}

\def\imath{i}
\def\jmath{j}

\def\feq{\ensuremath{f^{(\mathrm{eq})}}}
\def\eqref#1{(\ref{#1})}

\def\hatt{{\hat{0}}}
\def\hati{{\hat{\imath}}}

\def\halpha{{\hat{\alpha}}}
\def\hbeta{{\hat{\beta}}}
\def\hgamma{{\hat{\gamma}}}
\def\hrho{{\hat{\rho}}}

\def\omegabar{\overline{\omega}}

\def\Ki{{\rm Ki}}

\begin{document}

\title{Anderson-Witting transport coefficients for flows in general relativity}

\author{Victor E.~Ambru\cb{s}}

\affil{Department of Physics, West University of Timi\cb{s}oara,
Bd.~Vasile P\^arvan 4, Timi\cb{s}oara, RO 300223, Romania}
\corresp{victor.ambrus@e-uvt.ro}

\maketitle

\begin{abstract}
The transport coefficients induced by the Anderson-Witting approximation of the collision term 
in the relativistic Boltzmann equation are derived for close to equilibrium flows in general relativity.
Using the tetrad formalism, it is shown that the expression for these coefficients is the same as that 
obtained on flat space-time, in agreement with the generalized equivalence principle.
\end{abstract}

\section{INTRODUCTION}

Relativistic hydrodynamics constitutes a relatively new area of research, being fundamental 
for the understanding of fluid flows in extreme conditions, either when the typical velocities involved approach 
the speed of light, or when the space-time curvature becomes significant \cite{rezzolla13}. Applications from the 
first category include the quark-gluon plasma \cite{jacak12}, while from the second category, we mention astrophysical 
phenomena such as stellar collapse \cite{young04}, accretion problems \cite{banyuls97} or cosmology \cite{ellis12}.

In many problems of astrophysical importance where the 
flow is sufficiently rarefied that the hydrodynamic (continuum) approximation 
cannot be applied,
a kinetic theory description is required \cite{tsumura07}. Such an approach has the advantage that 
the set of hydrodynamic conservation equations, which is highly non-linear in the viscous 
regime, emerges from the relativistic Boltzmann equation, where the advection is performed 
in a simple manner. In particular, for flows not far from equilibrium, the hydrodynamic limit of the 
Boltzmann equation can be obtained through the Chapman-Enskog expansion \cite{cercignani02}. 

In this paper, we employ the Chapman-Enskog procedure to derive expressions for the transport coefficients 
when the Anderson-Witting approximation for the collision term in the relativistic Boltzmann equation is employed. 
We consider relativistic flows on an arbitrary background space-time, thus extending the 
results in Refs.~\cite{cercignani02,anderson74a}, obtained for the Minkowski space-time.
In our analysis, we highlight a procedure for obtaining expressions for the 
non-equilibrium contributions to the stress-energy tensor (SET), involving the computation of a special type 
of moments of the equilibrium distribution function, which we summarise in the appendix. The conclusion of 
our study is that the expressions for the transport coefficients is identical to those obtained on 
the flat Minkowski space-time in Refs.~\cite{cercignani02,anderson74a}. We also present a comparison 
with the results reported in Refs.~\cite{cercignani02,ambrus16} for the Marle model.

\section{ECKART AND LANDAU FRAMES}


In the Eckart frame \cite{eckart40}, the macroscopic velocity $u^\mu$ is defined to be parallel to the particle 
flow four-vector $N^\mu$, such that:
\begin{equation}
 N^\mu = nu^\mu, \qquad T^{\mu\nu} = E u^\mu u^\nu + (P + \omegabar) \Delta^{\mu\nu} + 2q^{(\mu} u^{\nu)} + \pi^{\mu\nu},
 \label{eq:eckart_tmunu}
\end{equation}
such that $u^\mu = N^\mu / \sqrt{-N_\mu N^\mu}$. 
In the above, $g^{\mu\nu}$ is the space-time metric, 
$\Delta^{\mu\nu} = u^\mu u^\nu + g^{\mu\nu}$ is the projector on 
the hypersurface perpendicular to $u^\mu$,
$n$ is the macroscopic number density, 
$E$ is the energy density, $P$ is the hydrostatic pressure, while the
dynamic pressure $\omegabar$, heat flux $q^\mu$ and pressure deviator $\pi^{\mu\nu}$ 
comprise the non-equilibrium terms.\footnote{We use the signature $(-,+,+,+)$ for the 
metric and geometrical units in which $c = G = 1$ throughout this paper.}

Once the SET and $u^\mu$ are known, all other quantities can be obtained using:
\begin{equation}
 n = -u_\mu N^\mu = \sqrt{-N_\mu N^\mu}, \qquad
 E = u_\mu u_\nu T^{\mu\nu}, \qquad 
 P + \omegabar = \frac{1}{3} \Delta_{\mu\nu} T^{\mu\nu},\qquad
 q^\mu = -\Delta^\mu{}_{\nu} u_{\lambda} T^{\nu\lambda}, \qquad
 \pi^{\mu\nu} = T^{<\mu\nu>},
 \label{eq:eckart_macro}
\end{equation}
where the notation $A^{<\mu\nu>}$ refers to:
\begin{equation}
 A^{<\mu\nu>} \equiv \left[\frac{1}{2}\left(\Delta^\mu{}_{\lambda} \Delta^\nu{}_{\sigma} 
 + \Delta^\mu{}_{\sigma} \Delta^\nu{}_{\lambda}\right) - 
 \frac{1}{3} \Delta^{\mu\nu} \Delta_{\lambda\sigma}\right] A^{\lambda\sigma}.
 \label{eq:angular_def}
\end{equation}

In the Landau (energy) frame \cite{landau87}, the four-velocity is defined as an eigenvector of $T^{\mu\nu}$:
\begin{equation}
 T^{\mu}{}_\nu u^\nu_L = -E_L u^\mu_L,\label{eq:landau_def}
\end{equation}
where the subscript $L$ indicates that the velocity and the energy density are expressed with respect to the 
Landau frame. Since $\Delta^\mu_{L;\nu} u_{L;\lambda} T^{\nu\lambda} = 0$, it can be seen that in this frame,
the heat flux (or energy dissipation) is everywhere nil, such that $N^\mu$ and $T^{\mu\nu}$ take the following form:
\begin{equation}
 N^\mu = n_L u^\mu_L + \mathcal{J}^\mu_L,\qquad 
 T^{\mu\nu} = E_L u^\mu_L u^\nu_L + (P_L + \omegabar_L) \Delta_L^{\mu\nu} + \pi^{\mu\nu}_L.
 \label{eq:landau_tmunu}
\end{equation}
Thus, in the Landau frame, $N^\mu$ and $u_L^\mu$ are no longer parallel. 
Close to local thermodynamic equilibrium, the quantity $\mathcal{J}^\mu_L$ can be linked to the heat 
flux of the Eckart frame via \cite{rezzolla13}:
\begin{equation}
 \mathcal{J}^\mu_L = -\frac{n}{E + P} q^\mu.\label{eq:landau_jmu}
\end{equation}

If the fluid is in local thermodynamic equilibrium, the Landau and Eckart frame coincide.
For small departures from equilibrium,
the nonequilibrium quantities $\omegabar$, $q^\mu$ and $\pi^{\mu\nu}$ can be written in 
terms of the thermodynamic forces $\nabla_\mu u^\mu$, 
$\Delta^{\mu\nu} \nabla_\nu T - u^\nu \nabla_\nu u^\mu$ and 
$\Delta^{\mu\nu} \nabla_{<\nu} u_{\mu>}$ as follows \cite{rezzolla13,cercignani02}:
\begin{equation}
 \omegabar = -\eta\nabla_{\mu} u^\mu, \qquad
 q^\mu= -\lambda \left(\Delta^{\mu\nu} \nabla_\nu T + T u^\nu \nabla_\nu u^\mu\right), \qquad 
 \pi_{\mu\nu} = -2\mu \nabla_{<\mu} u_{\nu>},\label{eq:tcoeff}
\end{equation}
which define the transport coefficients $\eta$, $\lambda$ and $\mu$, known as the 
coefficients of bulk viscosity, thermal conductivity and shear viscosity, respectively
\cite{cercignani02}. 

\section{BOLTZMANN EQUATION IN THE ANDERSON-WITTING APPROXIMATION}

The Boltzmann equation can be written in conservative form with respect to a tetrad field 
$e_\halpha^\mu$ 
as follows \cite{ambrus16,cardall13}:
\begin{equation}
 \frac{1}{\sqrt{-g}} \partial_\mu \left(\sqrt{-g} p^\halpha e_\halpha^\mu f\right) - 
 p^\hatt \frac{\partial}{\partial p^\hati} \left(
 \Gamma^\hati{}_{\halpha\hbeta} \frac{p^\halpha p^\hbeta}{p^\hatt} f\right) = J[f],
 \label{eq:boltz_cons}
\end{equation}
where hatted indices denote tetrad components, 
$p^\halpha$ represent the tetrad components of the on-shell 
particle four-momentum vector, 
$f$ is the one-particle distribution function 
and $J[f]$ represents the Boltzmann collision integral. 
The tetrad components of the hydrodynamic variables $N^\halpha$ and $T^{\halpha\hbeta}$ can be obtained 
as moments of $f$ \cite{cercignani02}:
\begin{equation}
 N^\halpha = \int \frac{d^3p}{p^\hatt} f\,p^\halpha, \qquad 
 T^{\halpha\hbeta} = \int \frac{d^3p}{p^\hatt} f\,p^\halpha p^\hbeta.
\end{equation}

%
Due to the complicated nature of the collision integral, model equations for the collision term are
customarily employed. 
Here, we consider the Anderson-Witting approximation \cite{cercignani02,anderson74a}:
\begin{equation}
 J_{\rm A-W}[f] = \frac{p \cdot u_L}{\tau}(f - \feq_L),
\end{equation}
where $\tau$ is the relaxation time. The equilibrium distribution function $\feq_L$ is defined 
in terms of quantities expressed with respect to the Landau frame:
\begin{equation}
 \feq_L = \frac{n_L}{4\pi m^2 T_L K_2(\zeta_L)} \exp\left(\frac{p\cdot u_L}{T_L}\right),
 \qquad N^{\halpha}_{\rm eq;L} = n_L u^\halpha_L, \qquad 
 T^{\halpha\hbeta}_{\rm eq; L} = E_L u^\halpha_L u^\hbeta_L + P_L \Delta^{\halpha\hbeta}_L.
 \label{eq:feq} 
\end{equation}
In the above, $m$ is the particle mass, $\zeta_L \equiv m / T_L$ is the relativistic coldness \cite{rezzolla13}, 
$K_n(\zeta_L)$ denotes the modified Bessel functions of the third kind, $P_L = n_L T_L$ is the hydrostatic pressure
and the Landau energy $E_L$ and Landau temperature $T_L$ are linked through $E_L = n_LmG(\zeta_L) - P_L$.

In Ref.~\cite{ambrus16}, the transport coefficients corresponding to the Marle collision term were 
analysed for flows in general relativity. In this paper, we extend the results of Ref.~\cite{ambrus16} 
by performing a similar analysis of the transport coefficients arising in the Anderson-Witting model. 

\section{CHAPMAN-ENSKOG EXPANSION}

In order to perform the Chapman-Enskog expansion, we consider $\delta f \equiv f - \feq$ and $\tau$ to 
be small. Thus, the deviation $\delta f$ from equilibrium can be approximated by keeping only
the zeroth order $f = \feq$ on the left hand side in Eq.~(\ref{eq:boltz_cons}), such that:
\begin{equation}
 \delta f = \tau\left[\frac{1}{\sqrt{-g}} \partial_\mu \left(\sqrt{-g} \frac{p^\halpha e_\halpha^\mu}{p\cdot u} \feq\right) - 
 p^\hatt \frac{\partial}{\partial p^\hati} \left(
 \Gamma^\hati{}_{\halpha\hbeta} \frac{p^\halpha p^\hbeta}{p^\hatt (p\cdot u)} \feq\right)
 + (\nabla_\halpha u_\hbeta) \frac{p^\halpha p^\hbeta}{(p\cdot u)^2} \feq\right].
 \label{eq:df}
\end{equation}
In the above, the subscript $L$ was dropped for quantities on the right hand side of the equation, since the 
Landau and the Eckart quantities coincide when $f = \feq$.
We emphasize that the above equation is different from the one obtained in Ref.~\cite{anderson74a},
since it is written in conservative form, such that its moments can 
be easily obtained.
Indeed, integrating the above equation over the momentum space gives:
\begin{equation}
 \delta T_{0}^{\halpha_1 \dots \alpha_n} = \tau\left[-\nabla_\hbeta T_{1}^{\hbeta\halpha_1 \dots \halpha_n} + 
 (\nabla_\hbeta u_\hgamma) T_{2}^{\hbeta\hgamma \halpha_1 \dots \halpha_n}\right],\quad 
 {\rm where} \quad 
 T_n^{\halpha_1 \dots \halpha_s} = \int \frac{d^3p}{p^\hatt} \, \frac{p^{\halpha_1} \cdots p^{\halpha_s}}{(-p\cdot u)^n} \feq,
 \label{eq:Tn_def}
\end{equation}
where the covariant derivative arises naturally due to the conservative form of Eq.~\eqref{eq:df}.
A comparison of the above equation with Eqs.(34) and (36) from Ref.~\cite{anderson74a} shows that,
besides moments of type $T^{\halpha_1 \dots \halpha_s}_n$ with $n = 1$, the conservative 
approach also requires the computation of the moments with $n = 2$. 
We present an analysis of such moments relevant for the present work in the Appendix.

\subsection{The coefficient of bulk viscosity $\eta$}

Taking the trace of Eq.~\eqref{eq:landau_tmunu} shows that $\delta T^\halpha{}_\halpha = 3\omegabar_L$, 
since $T^\halpha_{{\rm eq};\halpha} = -E_L + 3P_L$. Furthermore, the definition \eqref{eq:Tn_def} and 
the properties \eqref{eq:Tn_prop} can be used to show that $\delta T^\halpha{}_\halpha = -m^2 \delta T_{0}$.
Setting $n = 0$ in Eq.~\eqref{eq:Tn_def} gives:
\begin{equation}
 \omegabar = -\frac{\tau m^2}{3} \left[-\nabla_\hgamma T_1^\hgamma + (\nabla_\hgamma u_\hrho) T_2^{\hgamma \hrho}\right].
 \label{eq:omegabar_aux}
\end{equation}
The divergence of $T_1^\hgamma = T_{11}^0 u^\hgamma$ can be computed by substituting 
$T_{11}^0$ from Eq.~(\ref{eq:T1}):
\begin{equation}
 \nabla_\hgamma T^\hgamma_1 = T_{11}^0 \nabla_\hgamma u^\hgamma + D T_{11}^0 = 
 -\frac{P}{m^2} \left(1 - \frac{3}{c_v}\right) \nabla_\hgamma u^\hgamma,
 \label{eq:nabla_T110}
\end{equation}
where the notation $D = u^\hgamma \nabla_\hgamma$ was used. 
In the Chapman-Enskog procedure, the derivatives $D$ are replaced by making use of the 
conservation equations at the Euler level \cite{rezzolla13,ambrus16}: 
\begin{equation}
 Dn = -n \nabla_\hgamma u^\hgamma, \qquad 
 DE = -(E + P) \nabla_\hgamma u^\hgamma, \qquad 
 DT = - \frac{T}{c_v} \nabla_\hgamma u^\hgamma, \qquad 
 Du^\halpha = -\frac{1}{E + P} \Delta^{\halpha\hgamma} \nabla_\hgamma P,
 \label{eq:D}
\end{equation}
where the specific heat $c_v$ is given by:
\begin{equation}
 c_v = \zeta^2 + 5G \zeta - G^2 \zeta^2 - 1.
\end{equation}
In the above, $G\equiv G(\zeta) = K_3(\zeta) / K_2(\zeta)$.
Using Eqs.~\eqref{eq:Tcoeff_def} and Eq.~\eqref{eq:T2} for $T^{\hgamma\hrho}_2$, together 
with the property $u^\hrho \nabla_\hgamma u_\hrho = 0$, it can be shown that substituting 
Eq.~\eqref{eq:nabla_T110} into Eq.~\eqref{eq:omegabar_aux} allows $\eta$ to be put in the 
following form:
\begin{equation}
 \eta = \tau P \left(-1 - \frac{1}{c_v} + \frac{\zeta G}{3} - \frac{\zeta^2}{9} + \frac{\zeta^2 {\rm Ki}_2}{9K_2}\right),
 \label{eq:eta}
\end{equation}
where $\Ki_2 \equiv \Ki_2(\zeta)$ is defined in the Appendix.
It can be checked that the above result coincides with those obtained in Refs.~\cite{cercignani02,anderson74a} on 
flat space-time by using the following relation:
\begin{equation}
 {\rm Ki}_2(\zeta) = \zeta[-{\rm Ki}_1(\zeta) + K_1(\zeta)].\label{eq:ki2_ki1}
\end{equation}

\subsection{Coefficient of thermal conductivity $\lambda$}

In flows close to equilibrium, $\delta N^\halpha = -\frac{n}{P+E} q^\halpha$, 
by virtue of Eqs.~\eqref{eq:landau_tmunu} and \eqref{eq:landau_jmu}, since 
$N^\halpha_{\rm eq} = n_L u^\halpha_L$. Setting $n = 1$ in Eq.~\eqref{eq:Tn_def} 
allows the heat flux to be written as:
\begin{equation}
 q^\halpha = -\frac{\tau(P + E)}{n} \left[-\nabla_\hbeta T^{\halpha\hbeta}_1 + 
 (\nabla_\hbeta u_\hgamma) T^{\halpha\hbeta\hgamma}_2\right]
 = \frac{\tau(P + E)}{n} \left[n Du^\halpha + \Delta^{\halpha\hbeta} \nabla_\hbeta T^1_{12}\right],
 \label{eq:qheat_aux}
\end{equation}
where the second equality can be obtained by using the property 
$\Delta^\halpha{}_\hbeta q^\hbeta$= $q^\halpha$.
The derivative of $T_{12}^1$ can be obtained from Eqs.~\eqref{eq:Tcoeff_def} and \eqref{eq:T1}:
\begin{equation}
 \nabla_\hbeta T^1_{12} = \frac{1}{3m}\left[5\zeta - \zeta^2 G(\zeta) + 
 \frac{\zeta^2 {\rm Ki}_1(\zeta)}{K_2(\zeta)}\right] \nabla_\hbeta P + 
 \frac{P}{3m^2} \left[3\zeta^2 - 5\zeta^3 G(\zeta) + \zeta^4 G^2(\zeta) - \frac{\zeta^4 G(\zeta) {\rm Ki}_1(\zeta)}{K_2(\zeta)}\right]
 \nabla_\hbeta T.
\end{equation}
Using Eq.~\eqref{eq:D} to replace $Du^\halpha$ in the expression for $q^\halpha$ in Eq.~\eqref{eq:tcoeff}, as well as in 
Eq.~\eqref{eq:qheat_aux}, it can be shown that the thermal conductivity is given by:
\begin{equation}
 \lambda = \tau P \frac{\zeta^4 G(\zeta)}{3m} \left[\frac{G(\zeta) {\rm Ki}_1(\zeta)}{K_2(\zeta)} - 
 G^2(\zeta) + \frac{5G(\zeta)}{\zeta} - \frac{3}{\zeta^2}\right].
\end{equation}
This result is in agreement with the one reported in Refs.~\cite{cercignani02,anderson74a}.

\subsection{Coefficient of shear viscosity $\mu$}

Finally, $\mu$ can be obtained from the expression for $\pi^{\halpha\hbeta}$ by using Eq.~\eqref{eq:tcoeff}.
By noting that $\delta T^{\halpha\hbeta}_0 = \omegabar \Delta^{\halpha\hbeta} + \pi^{\halpha\hbeta}$, 
it can be shown that:
\begin{equation}
 \pi^{\halpha\hbeta} = \eta \Delta^{\halpha\hbeta} \nabla_\hgamma u^\hgamma + 
 \tau\left[-\nabla_\hgamma T^{\halpha\hbeta\hgamma}_1 + 
 (\nabla_\hgamma u_\hrho) T^{\halpha\hbeta\hgamma\hrho}_2\right].
 \label{eq:pi_aux}
\end{equation}
Using Eqs.~\eqref{eq:Tcoeff_def} and \eqref{eq:T1} for $T^{\halpha\hbeta\hgamma}_1$, 
the following result can be obtained:
\begin{equation}
 \nabla_\hgamma T^{\halpha\hbeta\hgamma}_1 =
 nT \nabla_\hgamma u^\hgamma \left[\left(\frac{5}{3} - \frac{1}{c_v}\right) \Delta^{\halpha\hbeta} - \eta^{\halpha\hbeta}\right] + 
 nT D(u^\halpha u^\hbeta) + 2nT \nabla^{<\halpha} u^{\hbeta>}.
\end{equation}
By susbtituting the above result in Eq.~\eqref{eq:pi_aux} and using Eqs.~\eqref{eq:Tcoeff_def} and \eqref{eq:T2} for 
$T^{\halpha\hbeta\hgamma\hrho}$, $\pi^{\halpha\hbeta}$ can be written as:
\begin{equation}
 \pi^{\halpha\hbeta} = \tau n T\left[\Delta^{\halpha\hbeta} \nabla_\hgamma u^\hgamma \left(\frac{\eta}{\tau n T} + \frac{5 T_{24}^1}{3nT} - 
 \frac{7}{3} + \frac{1}{c_v} \right) + 2\left(\frac{T_{24}^1}{nT} - 2\right) \nabla^{<\halpha} u^{\hbeta>}\right].
\end{equation}
Using the expression \eqref{eq:eta} obtained for $\eta$, it can be shown that the coefficient 
of $\nabla_\hgamma u^\hgamma$ vanishes, such that:
\begin{equation}
 \mu = \frac{3}{5}\left(\eta + \tau n T \frac{1+c_v}{c_v}\right) = 
 \frac{\tau n T}{15} \zeta 
 \left(3G(\zeta) - \zeta + \frac{\zeta {\rm Ki}_2(\zeta)}{K_2(\zeta)}\right).
 \label{eq:mu}
\end{equation}
Employing Eq.~\eqref{eq:ki2_ki1} shows that Eq.~\eqref{eq:mu} reduces to the result in Refs.~\cite{cercignani02,anderson74a}.

\subsection{Analysis of the results and comparison to the Marle model}

\begin{figure}
\begin{tabular}{cc}
 \includegraphics[width=0.4\linewidth]{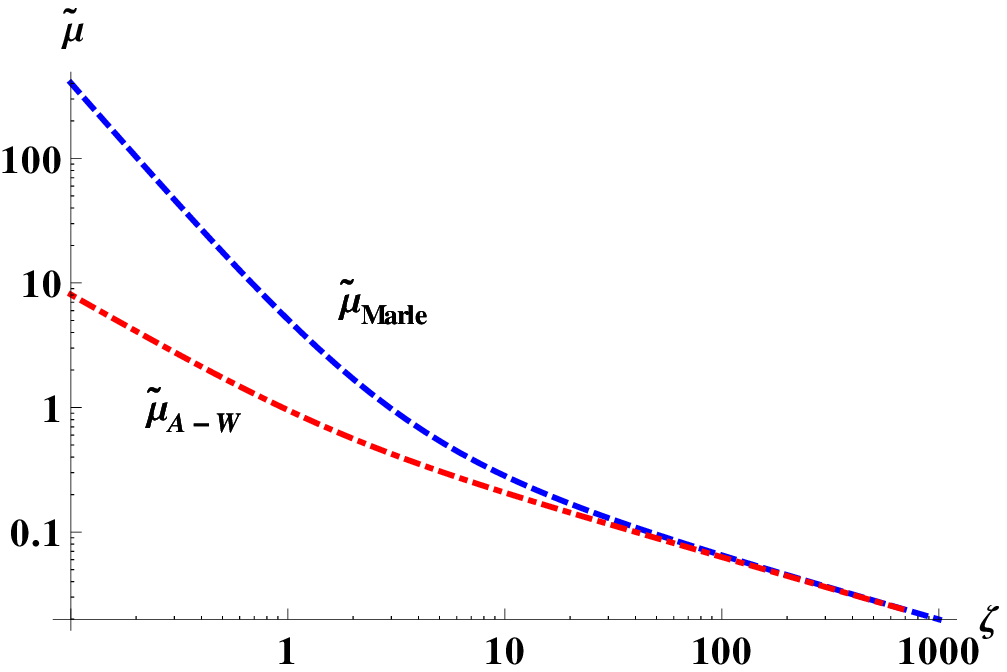} & 
 \includegraphics[width=0.4\linewidth]{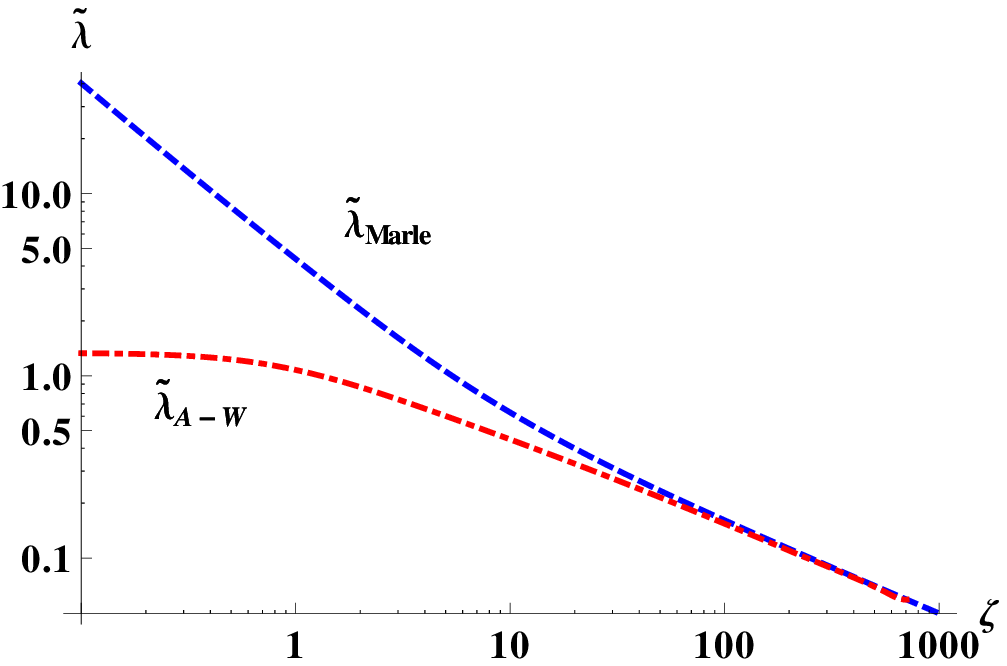} \\ 
 \includegraphics[width=0.4\linewidth]{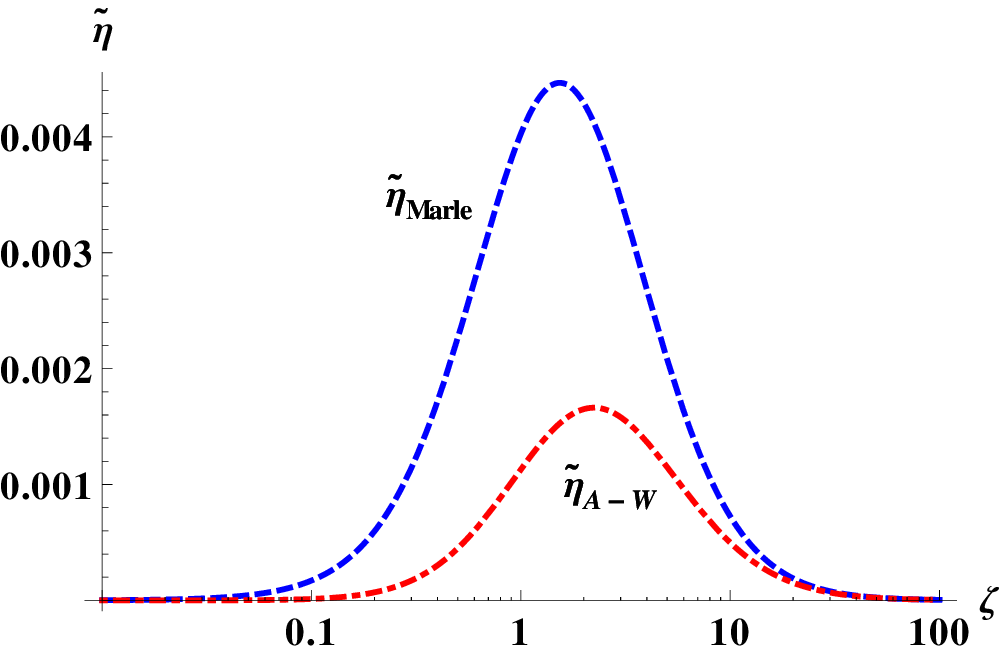} & 
 \includegraphics[width=0.4\linewidth]{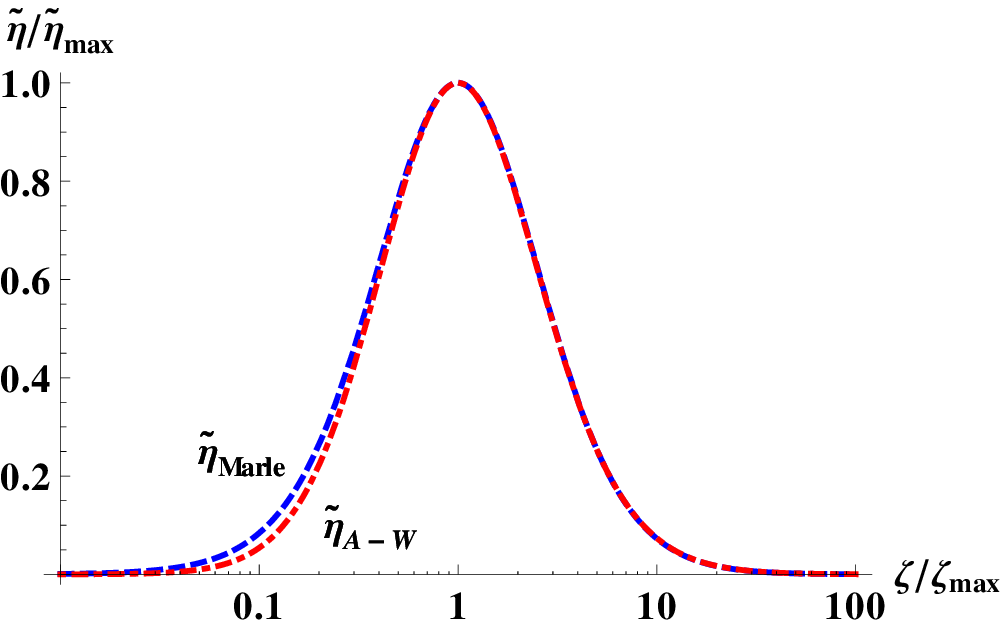}
\end{tabular}
\caption{Comparison of the (a) shear viscosity, (b) thermal conductivity and 
(c) bulk viscosity coefficients obtained when the Anderson-Witting and Marle approximations 
for the collision term are employed. (d) The curves corresponding to $\eta_{\rm M}$ and $\eta_{\rm A-W}$ are 
almost overlapped after the scaling in Eq.~\eqref{eq:eta_scaled}.}
\label{fig:aw_vs_m}
\end{figure}

In this section, the transport coefficients obtained here for the Anderson-Witting model are compared to those obtained 
using the Marle model when the relaxation time is set to \cite{cercignani02,ambrus16}:
\begin{equation}
 \tau = \frac{1}{n \sigma \braket{v}},\qquad 
 \braket{v} = \frac{\zeta_E}{K_1(\zeta_E)} \left[e^{-\zeta_E} \frac{1+\zeta_E}{\zeta_E^2} - 
 \Gamma(0, \zeta_E)\right],\label{eq:tau}
\end{equation}
where $\sigma$ is the differential cross section and $\Gamma(\nu, z)$ is the incomplete 
Gamma function \cite{olver10}.
In order to perform the comparison, ``effective'' transport coefficients may be defined, such 
that they only depend on $\zeta = m / T$ \cite{ambrus16}:
\begin{equation}
 \widetilde{\eta} = \frac{\sigma \eta}{m}, \qquad 
 \widetilde{\lambda} = \sigma \lambda, \qquad 
 \widetilde{\mu} = \frac{\sigma \mu}{m}.
 \label{eq:tcoeff_eff}
\end{equation}
The log-log plots in Fig.~\ref{fig:aw_vs_m}(a) and (b) show that the behaviour of $\widetilde{\mu}$ and 
$\widetilde{\lambda}$ in the non-relativistic (large $\zeta$) limit is the same in the Marle and Anderson-Witting models.
However, as $\zeta\rightarrow 0$, the effective heat conductivity $\widetilde{\lambda}_{\rm A-W}$ in the Anderson-Witting 
model tends to $4/3$, while in the Marle model, $\widetilde{\lambda}_{\rm M} \rightarrow \infty$. Even though not visible in the 
plot, $\widetilde{\lambda}_{\rm A-W}$ has a point of local maximum at $\zeta = 0.023151$, where its value differs only 
slightly from its value at $\zeta = 0$ ($\widetilde{\lambda}_{\rm max} = 1.33348$). 
In the case of the coefficient of bulk viscosity, Fig.~\ref{fig:aw_vs_m}(c) 
shows that it exhibits the same main features in both models, namely: it decreases to $0$ as $\zeta \rightarrow 0$ and 
$\zeta \rightarrow \infty$ and it attains a maximum value at a finite value of $\zeta$. While in the Marle model, the 
maximum is attained at $\zeta_{\rm max;M} = 1.535$, where $\widetilde{\eta}_M(\zeta_{\rm max; M}) = 0.0044664$, 
in the A-W model, $\zeta_{\rm max; A-W} = 2.23578$ and $\widetilde{\eta}_M(\zeta_{\rm max; M}) = 0.00166279$. 
It is surprising to note that the bulk viscosity in the two models are almost exactly related through a scaling 
of the argument and overall value such that their point of maximum coincides, i.e.:
\begin{equation}
 \frac{\widetilde{\eta}_{\rm M}\left(\zeta/\zeta_{\rm max; M}\right)}{\widetilde{\eta}_{\rm M}(\zeta_{\rm max; M})} \simeq 
 \frac{\widetilde{\eta}_{\rm A-W}\left(\zeta/\zeta_{\rm max; A-W}\right)}{\widetilde{\eta}_{\rm A-W}(\zeta_{\rm max; A-W})}.
 \label{eq:eta_scaled}
\end{equation}
Fgure~\ref{fig:aw_vs_m}(d) shows a comparison between the two sides of the above equation.

\vspace{-10pt}
\section{CONCLUSION}
In this paper, we employed the conservative form of the Boltzmann equation based on the tetrad formalism 
to obtain expressions for the transport coefficients corresponding to the Anderson-Witting model for 
reltivistic flows in general relativity. Our conclusion is that the form of these coefficients coincides 
with that obtained in flat space-time, in agreement with the equilvalence principle. Further, a graphical 
analysis showed that an appropriate scaling brings the coefficient of bulk viscosity $\eta$ in the 
Anderson-Witting to a form which is very close to that in the Marle model.
\section{ACKNOWLEDGMENTS}
This work was supported by a grant of the Romanian National Authority for Scientific Research and Innovation, CNCS-UEFISCDI, project number 
PN-II-RU-TE-2014-4-2910.

\appendix
\section{APPENDIX}\label{app:T}

In this section, the moments of $\feq$ defined in Eq.~\eqref{eq:Tn_def} are computed 
for $n = 0, 1, 2$ with $s$ ranging from $0$ up to $2$, $3$ and $4$, respectively.
In order to compute these moments, the following notation is employed:
\begin{eqnarray}
 T_n = T_{n0}^0, \qquad 
 T_n^{\halpha} = T_{n1}^0 u^\halpha, \qquad 
 T_n^{\halpha\hbeta} = T_{n2}^0 u^\halpha u^\hbeta + T_{n2}^1 \eta^{\halpha\hbeta}, \qquad 
 T_n^{\halpha\hbeta\hgamma} = T_{n3}^0 u^\halpha u^\hbeta u^\hgamma + T_{n3}^1 u_\hrho \Delta^{\halpha\hbeta\hgamma\hrho}, \nonumber\\
 T_n^{\halpha\hbeta\hgamma\hrho} = T_{n4}^0 u^\halpha u^\hbeta u^\hgamma u^\hrho + 
 T_{n4}^1 (u^\halpha u^\hbeta \eta^{\hgamma\hrho} + u^\halpha u^\hgamma \eta^{\hbeta\hrho} + 
 u^\halpha u^\hrho \eta^{\hbeta\hgamma} + u^\hbeta u^\hgamma \eta^{\halpha\hrho} + 
 u^\hbeta u^\hrho \eta^{\halpha\hgamma} + u^\hgamma u^\hrho \eta^{\halpha\hbeta}) +
 T_{n4}^2 \Delta^{\halpha\hbeta\hgamma\hrho},
 \label{eq:Tcoeff_def}
\end{eqnarray}
where $\Delta^{\halpha\hbeta\hgamma\hrho} = \eta^{\halpha\hbeta} \eta^{\hgamma\hrho} + 
\eta^{\halpha\hgamma} \eta^{\hbeta\hrho} + \eta^{\halpha\hrho} \eta^{\hbeta\hgamma}$.
Finding analytic expressions for the coefficients $T_{ns}^p$ defined in Eqs.~\eqref{eq:Tcoeff_def} 
makes the subject of the present section.

%
For the case $n = 0$, the relevant coefficients for the moments up to $s \le 2$ are:
\begin{equation}
 T_{00}^0 = \frac{n}{m}\left(G - \frac{4}{\zeta}\right), \qquad 
 T_{01}^0 = n, \qquad 
 T_{02}^0 = nm G, \qquad 
 T_{02}^1 = \frac{n m}{\zeta},
 \label{eq:T0}
\end{equation}
where $G \equiv G(\zeta) = K_3(\zeta) / K_2(\zeta)$.

The relevant coefficients for the case when $n = 1$ are:
\begin{eqnarray}
 T_{10}^0 = \frac{n}{m^2}\left(\zeta G - 4 - \frac{\zeta \Ki_1}{K_2}\right), \qquad 
 T_{11}^0 = \frac{n}{m} \left(G - \frac{4}{\zeta}\right), \qquad 
 T_{12}^0 = \frac{n}{3}\left(8 - \zeta G + \frac{\zeta \Ki_1}{K_2}\right), \nonumber\\ 
 T_{12}^1 = \frac{n}{3}\left(5 - \zeta G + \frac{\zeta \Ki_1}{K_2}\right), \qquad
 T_{13}^0 = nm \left(G + \frac{2}{\zeta}\right), \qquad 
 T_{13}^1 = \frac{nm}{\zeta},
 \label{eq:T1}
\end{eqnarray}
where it is understood that the argument of $G$ and of all modified Bessel functions is $\zeta$, while 
$\Ki_n \equiv \Ki_n(\zeta) = \int_0^\infty dt \, (\cosh t)^{-n} e^{-\zeta \cosh t}$ denotes the integral 
of the modified Bessel functions \cite{cercignani02}.
The results in Eq.~\eqref{eq:T1} for $T_{12}^0$, $T_{12}^1$, $T_{13}^0$ and $T_{13}^1$ 
are in exact agreement with those reported in Eqs.~(44)-(47) of Ref.~\cite{anderson74a}.

Finally, the coefficients for the case when $n = 2$ are given by:
\begin{eqnarray}
 T_{20}^0 = \frac{n}{m^3} \left(\zeta +\frac{8}{\zeta} - 2G - \frac{\zeta \Ki_2}{K_2}\right), \qquad 
 T_{21}^0 = T_{10}^0, \qquad 
 T_{22}^0 = \frac{n}{3m}\left(6G - \frac{24}{\zeta} - \zeta + \frac{\zeta \Ki_2}{K_2}\right), \quad \nonumber\\
 T_{22}^1 = \frac{n}{3m}\left(3G - \frac{12}{\zeta} - \zeta + \frac{\zeta \Ki_2}{K_2}\right), \qquad 
 T_{23}^0 = n\left(6 - \zeta G + \frac{\zeta \Ki_1}{K_2}\right), \qquad 
 T_{23}^1 = T_{12}^1, \qquad \quad\nonumber\\
 T_{24}^0 = \frac{nm}{5} \left(2G + \frac{40}{\zeta} + \zeta - \frac{\zeta\Ki_2}{K_2}\right),\qquad
 T_{24}^1 = \frac{nm}{15} \left(-3G + \frac{30}{\zeta} + \zeta - \frac{\zeta\Ki_2}{K_2}\right), \qquad
 T_{24}^2 = T_{24}^1 + nT.
 \label{eq:T2}
\end{eqnarray}
As before, all functions depend on $\zeta$, unless otherwise indicated.

Before ending this Appendix, it is worth mentioning that the moments $T^{\halpha_1\dots\halpha_s}_n$ 
satisfy the following relations:
\begin{equation}
 -u_{\halpha_s} T_n^{\halpha_1 \dots \halpha_s} = T_{n-1}^{\halpha_1 \dots \halpha_{s-1}}, \qquad 
 -\eta_{\halpha_{s - 1} \halpha_{s}} T_n^{\halpha_1 \dots \halpha_s} = m^2 T_n^{\halpha_1 \dots \halpha_{s-2}}. 
 \label{eq:Tn_prop}
\end{equation}



\bibliographystyle{aipnum-cp}%
\bibliography{rotboltz-tim15}%

\end{document}